\documentclass{article}
\usepackage{LaThuileFPSpro}
\usepackage{graphicx}

\newcommand {\lapprox}
   {\raisebox{-0.7ex}{$\stackrel {\textstyle<}{\sim}$}}
\def\gsim{\,\lower.25ex\hbox{$\scriptstyle\sim$}\kern-1.30ex%
\raise 0.55ex\hbox{$\scriptstyle >$}\,}
\def\lsim{\,\lower.25ex\hbox{$\scriptstyle\sim$}\kern-1.30ex%
\raise 0.55ex\hbox{$\scriptstyle <$}\,}

\newcommand{\gev}{\ensuremath{\mathrm{GeV}}}

\newcommand{\MET}{\mbox{\ensuremath{E \kern-0.6em\slash_{\rm T}}}}
\newcommand{\MHT}{\mbox{\ensuremath{H \kern-0.75em\slash_{\rm T}}}}

\newcommand{\ra}{\ensuremath{\rightarrow}}

\newcommand{\pbi}{\ensuremath{\mathrm{pb}^{-1}}}
\newcommand{\fbi}{\ensuremath{\mathrm{fb}^{-1}}}

\begin{document}
\title{ 
  SEARCHES FOR THE HIGGS BOSON AND SUPERSYMMETRY
  AT THE TEVATRON
  }
\author{
  Thomas Nunnemann \\
  {\em Department f\"{u}r Physik, Ludwig-Maximillians Universit\"{a}t M\"{u}nchen}\\ 
  {\em Am Coulombwall 1, D-85748 Garching, Germany} \\
  On behalf of the D0 and CDF collaborations
  }
\maketitle

\baselineskip=11.6pt

\begin{abstract}
  The D0 and CDF experiments at the proton-antiproton collider Tevatron
  have extensively searched for the Higgs boson and signals of supersymmetry
  using a wide range of signatures.
  The status of these searches is reviewed with a focus on recent 
  measurements.
\end{abstract}
\newpage
\section{Introduction}
At the Tevatron collider, one of the main challenges is the search for the 
Higgs boson and for supersymmetric particles. The high integrated luminosities
being collected by both the CDF and D0 experiments enable searches with
unprecedented sensitivity. At the beginning of 2007, both experiments have
recorded data sets of more than 2\,\fbi. Recent results obtained with up 
to 1.1\,\fbi{} are presented in this note. All limits quoted are at 95\%
confidence level.

\section{Searches for the standard model Higgs boson}
In the standard model (SM) the Higgs mechanism is responsible for the 
electroweak symmetry breaking, thereby generating the masses of the $Z$ and
$W$ bosons. As a consequence of this mechanism a single neutral scalar
particle, namely the Higgs boson, remains after the symmetry breaking.
Assuming the validity of the standard model, global fits to
the electroweak data prefer a relatively low mass for the Higgs boson,
$m_H = 85^{+39}_{-28}\,\gev$\cite{Alcaraz:2006mx}, while direct
searches at the LEP collider set a lower bound on the mass of
$114.4\,\gev$\cite{Barate:2003sz}.

At low masses, $m_H \lapprox 135\,\gev$, the SM Higgs boson
dominantly decays via $H\rightarrow b\bar{b}$. For the main
production channel, which is the gluon-gluon fusion process 
$gg\rightarrow H$ this
leads to signatures which are irreducible from QCD production of
$b\bar{b}$ pairs. 
Therefore, at the Tevatron the highest sensitivity for low mass
Higgs bosons is obtained from the associated production of the Higgs boson
with the weak bosons, i.e. $WH$ and $ZH$.
At high masses the SM Higgs boson predominantly decays into $WW$ boson pairs,
which has a manageable background for the $gg\rightarrow H$ production mode.

\subsection{Low-mass Higgs boson, $m_H \lsim 135\,\gev$}
Both the CDF and D0 collaborations searched for low mass Higgs bosons using
the $WH\ra \ell\nu b\bar{b}$, $ZH\ra \nu\bar{\nu} b\bar{b}$, and 
$ZH\ra \ell\ell b\bar{b}$ production and decay modes. Dominating backgrounds
in these searches are the associated production of the weak bosons with 
$b\bar{b}$ pairs, $Wb\bar{b}$ and $Zb\bar{b}$, as well as the associated 
production $Wjj$ and $Zjj$ with jets originating from light-flavor 
quarks, which are falsely identified as $b$-jets.

\begin{figure}
  \begin{center}
  \includegraphics[width=0.38\textwidth]{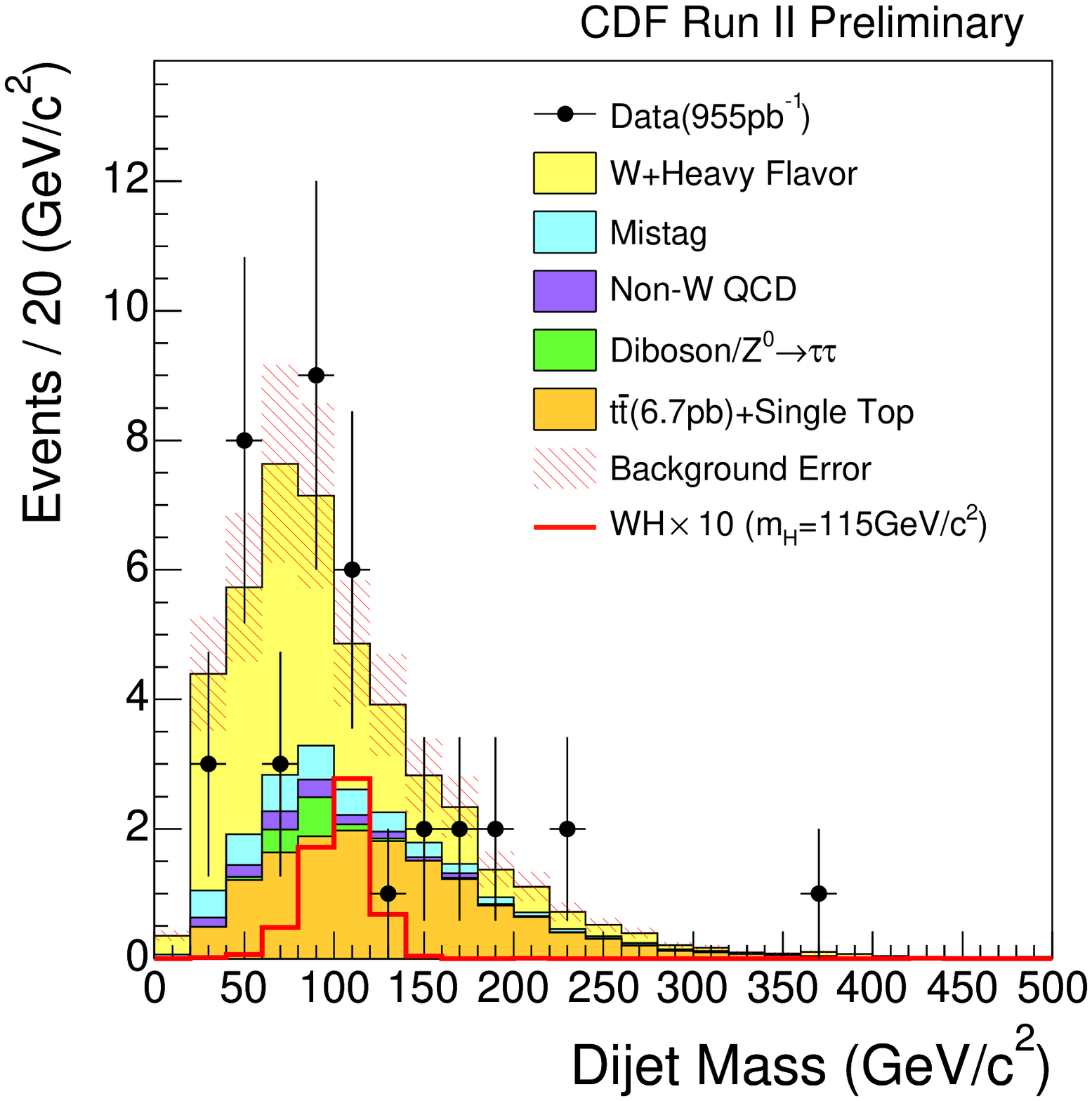}
  \includegraphics[width=0.57\textwidth,bb=0 0 567 355,clip]{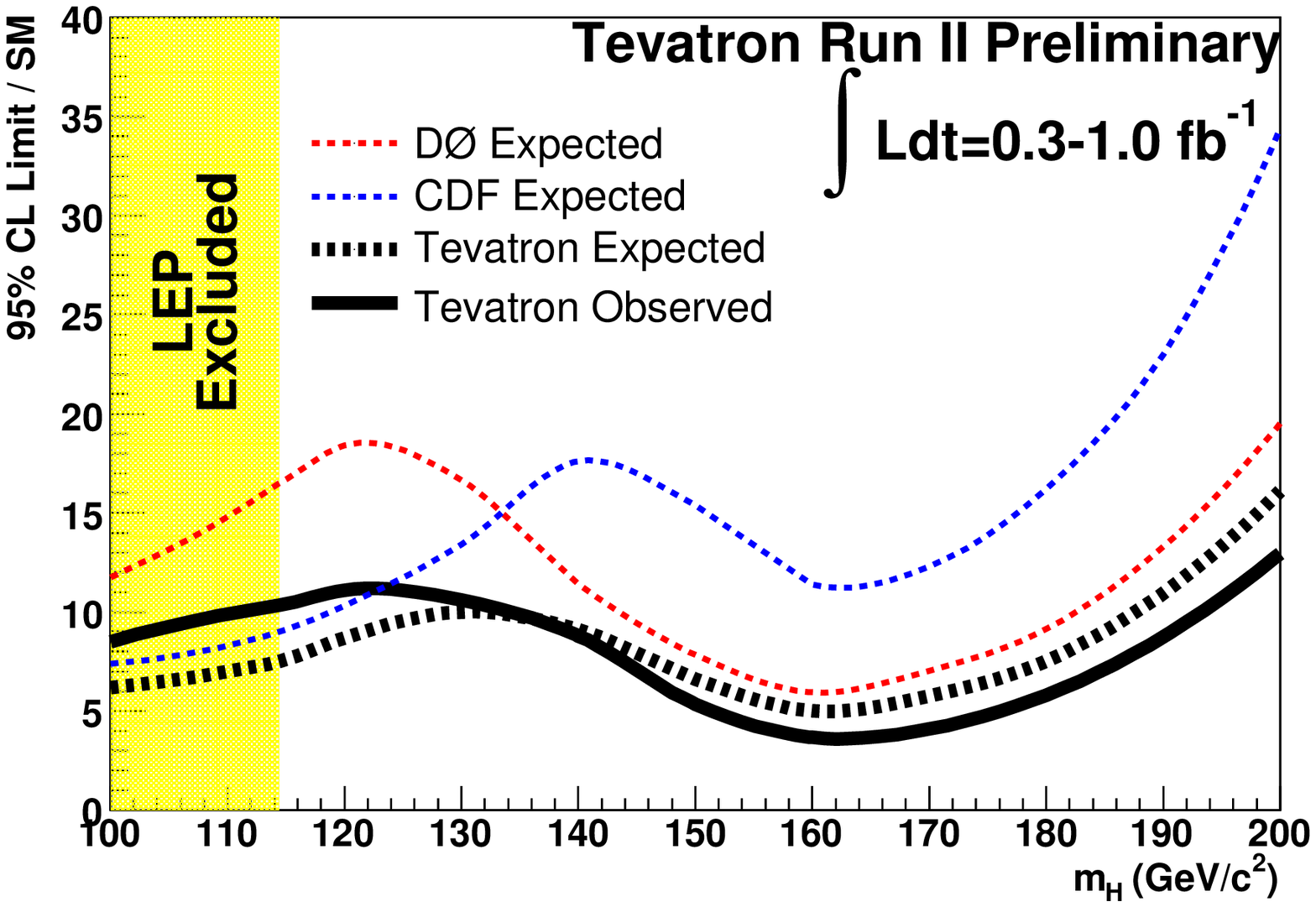}
  \caption{\it Left: dijet mass distribution in $W+\ge 2$ jets events with
    two secondary vertex tags. Right: combined CDF and D0 limits on SM Higgs 
    production, normalized to the predicted SM cross section.
    \label{fig:higgs}}
  \end{center}
\end{figure}

The CDF collaboration recently presented a search for the Higgs boson in
$WH\ra \ell\nu b\bar{b}$ production based on an integrated luminosity of 
1\,\fbi\cite{cdf8390}.
The event selection required a reconstructed electron or muon with a
transverse momentum $p_T>20\,\gev$, two jets with transverse energy 
$E_T>15\,\gev$ and large missing transverse momentum $\MET > 20\,\gev$.
The jets were identified to originate from $b$ quarks using secondary vertex
(SV) and neural network (NN) tagging algorithms. A resonant peak in the
dijet mass distribution, $M_{jj}$, indicative of $H\ra b\bar{b}$ was searched
for. The $M_{jj}$ distribution for events with two heavy-flavor jets 
identified using the
SV tagger is shown in Fig.~\ref{fig:higgs} together with the background
prediction and the expected Higgs signal. Upper limits on the production
cross sections, $\sigma_{95}$, were derived as function of Higgs boson mass 
$m_H$.
For $m_H \sim 115\,\gev$ the cross section limit from this measurement alone
compared to the SM prediction, $\sigma_{SM}$, corresponds to a sensitivity
of $\sigma_{95}/\sigma_{SM}\sim 20$.

The search for $ZH\ra \nu\bar{\nu} b\bar{b}$ production
has also a notable sensitivity
to $WH\ra \ell\nu b\bar{b}$ as the lepton might be undetected. Based on a data
sample of 1\,\fbi, the CDF collaboration searched for the Higgs boson in
events with large $\MET$ and two jets, of which one was required to be 
tagged\cite{cdf8442}.
In addition to $Zjj$, a large background contribution was found to be due to
QCD multijet production. For $m_H\sim 115\,\gev$ a sensitivity of 
$\sigma_{95}/\sigma_{SM}\sim 30$ was separately obtained for $ZH$ and 
$WH$ production. Combining both production modes the sensitivity was
$\sigma_{95}/\sigma_{SM}\sim 16$.

The $ZH\ra \ell\ell b\bar{b}$ channel is disfavored due to the low 
$Z\ra \ell\ell$ branching fraction. Nevertheless, the clear event topology
provides good background separation. The D0 collaboration recently presented
a search in this channel based on an integrated luminosity of 
0.9\,\fbi\cite{d05275}.
The analysis required a reconstructed $ee$ or $\mu\mu$ pair with a dilepton 
mass
consistent with the $Z$ boson mass and at least two jets which were required to
be identified as $b$ jets using the NN tagger. For central 
pseudorapidities, $|\eta| < 1.5$, a $b$-tagging efficiency of 72\% at a 
light-jet fake rate of 4\% was obtained. This search and a similar
CDF measurement\cite{cdf8422} were found to have sensitivities 
$\sigma_{95}/\sigma_{SM}\sim 25-30$ at $m_H\sim 115\,\gev$.

\subsection{High-mass Higgs boson, $m_H \gsim 135\,\gev$}
The dominant decay mode for $m_H \gsim 135\,\gev$ is $H\ra WW^{(*)}$. 
The $W$ decays into an electron or muon are used to suppress the QCD multijet
background. As the Higgs boson has spin-0, the final-state leptons are
predominately produced with small azimuthal separation due to spin-correlations 
between them. Therefore, the Higgs signal can be discriminated from the
electroweak production of $WW$ boson pairs.

The D0 collaboration performed a preliminary search based on 0.95\,\fbi{} using
the $ee$, $e\mu$, and $\mu\mu$ final states\cite{d05063_5194}. 
At $m_H\sim 160\,\gev$, where this
channel has optimal sensitivity, a cross-section ratio 
$\sigma_{95}/\sigma_{SM}\sim 4$ was obtained, which excludes models with four
fermion families\cite{Arik:2001iw} for $m_H \sim 150-185\,\gev$.

\subsection{Combined limits on Higgs boson production}
The CDF and D0 limits on SM Higgs production were combined for the first time
in summer 2006\cite{cdf8384_d05227}. Fig.~\ref{fig:higgs} shows the cross-section ratio
$\sigma_{95}/\sigma_{SM}$ as function of assumed Higgs boson mass $m_H$.
The combination does not yet include all searches presented above. After the
conference a new combination with significantly improved cross-section limits
was obtained, which also includes additional results obtained since then.

\section{Searches for neutral supersymmetric Higgs bosons}
Models with two Higgs-doublets, such as the minimal supersymmetric extension
of the standard model (MSSM), predict five physical Higgs bosons, of which
three ($h$, $H$, $A$) have neutral electric charge.
The phenomenology at large $\tan\beta$ (the ratio of the Higgs vacuum 
expectation values) is remarkable: The cross section for the gluon-gluon
fusion process $gg\ra H$ and the associated production $b\bar{b}H$ is 
largely enhanced and the $CP$-odd $A$ boson is nearly mass-degenerate 
with either the light or heavy $CP$ even state, $h$ or $H$, respectively.
The leading decay modes of the two mass-degenerate states, both denoted as 
$\phi$, are $\phi\ra b\bar{b}$ ($\sim 90\%$) and 
$\phi\ra \tau\tau$ ($\sim 10\%$). Despite the smaller branching fraction,
Higgs searches in the di-$\tau$ channel have the advantage of a much smaller
background level from multi-jet production.

\subsection{Supersymmetric Higgs in multi-jet events: $b\bar{b}\phi\ra b\bar{b}b\bar{b}$}
The D0 collaboration searched for the supersymmetric Higgs boson in the channel
$b\bar{b}\phi\ra b\bar{b}b\bar{b}$ using the dijet mass distribution
in events with three identified heavy-flavor jets.
The published analysis\cite{Abazov:2005yr} based on an integrated luminosity
of 260\,\pbi{} excludes a region at high $\tan\beta$, e.g. for 
$m_A~\sim 120\,\gev$ the constraint on $\tan\beta$ is $\tan\beta \lsim 50-60$ 
(depending on the assumed mixing in
the scalar top quark sector).
The preliminary update based on 0.9\,\fbi{} found exclusion limits improved by
about a third\cite{d05503}.

\subsection{Supersymmetric Higgs decaying to tau pairs:
$\phi\ra \tau\tau$}
\begin{figure}
  \begin{center}
  \includegraphics[width=0.58\textwidth]{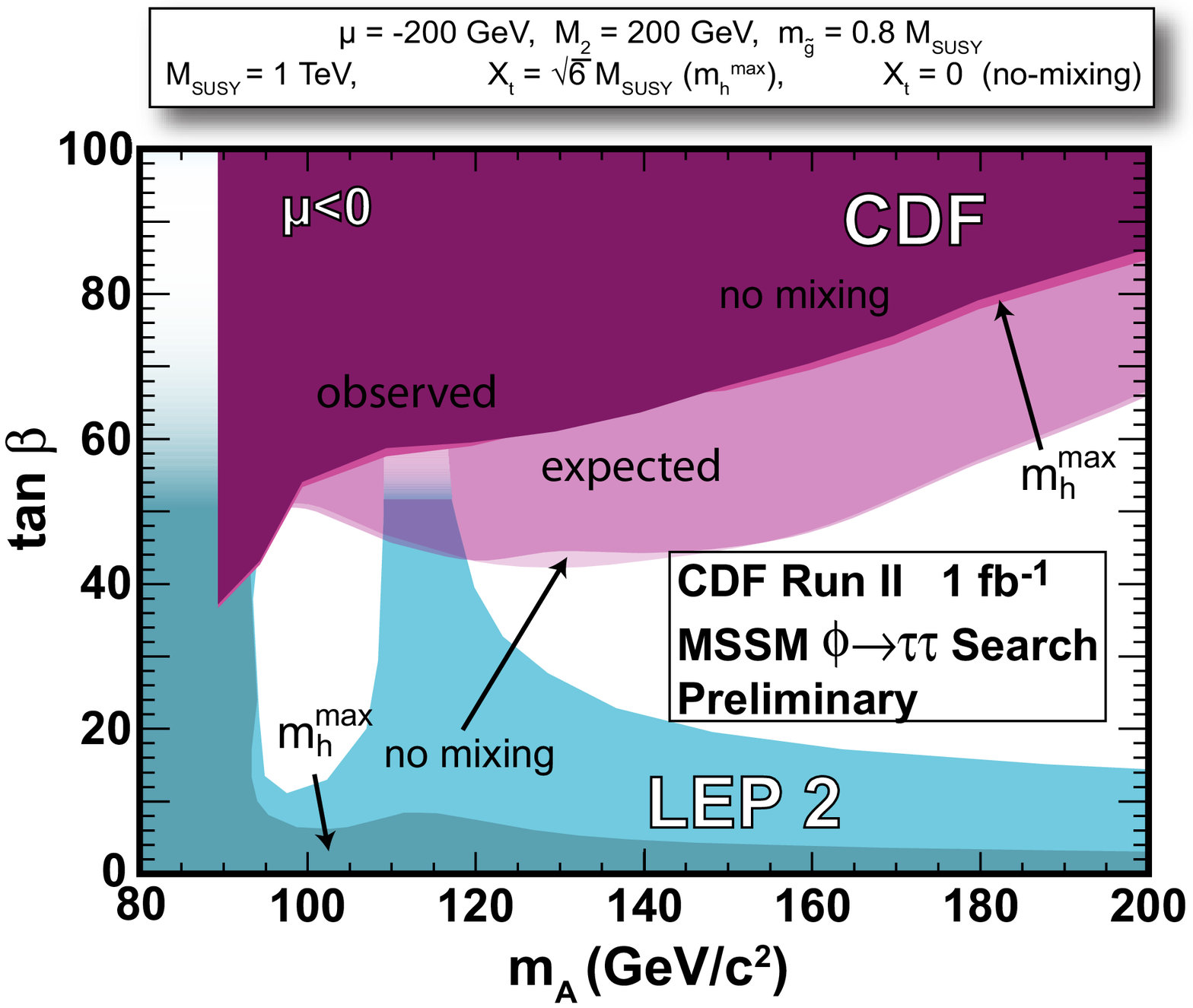}
  \includegraphics[width=0.39\textwidth,bb=35 32 210 275, clip]{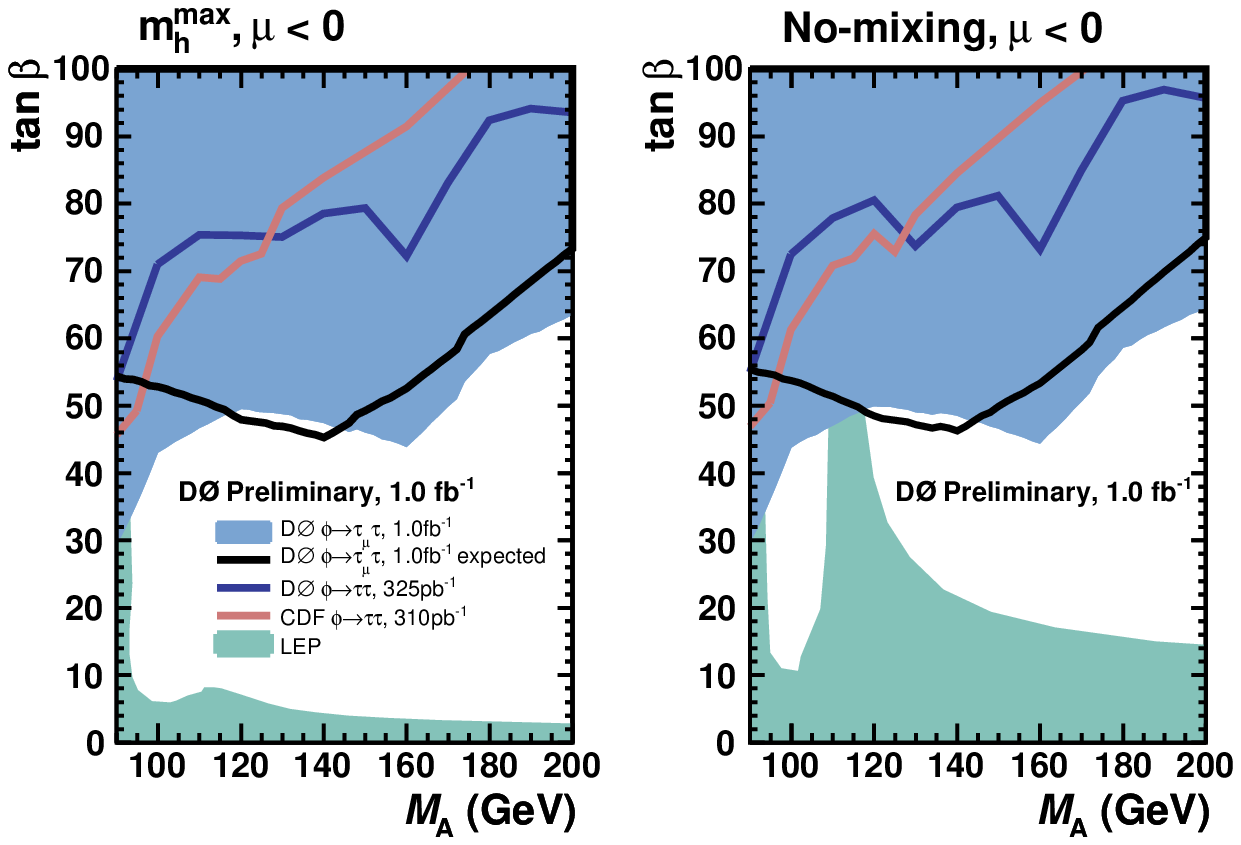}
  \caption{\it Excluded regions in the $tan\beta - m_A$ plane obtained by the
    CDF and D0 collaborations.
    \label{fig:mssmhiggs}}
  \end{center}
\end{figure}
Both, the CDF and D0 collaborations searched for the MSSM Higgs boson
decaying via $\phi\ra \tau\tau$ using data samples of 1\,\fbi{} each.
Whereas the CDF collaboration analyzed $\tau$-decays leading to $e\mu$,
$e\tau_h$, and $\mu\tau_h$ final states\cite{cdf8676} (with $\tau_h$ denoting hadronically
decaying $\tau$'s), the D0 selection\cite{d05331} required one $\tau$ decaying
into a muon.
The CDF collaboration observed a small excess of events ($<2\sigma$, only 
$e\tau_h$ and $\mu\tau_h$ channels) in
the visible mass distribution, which approximates the mass of the hypothetical
di-$\tau$ resonance. This non-significant excess was not confirmed by the
D0 search. The exclusion regions in the plane given by $m_A$ and $\tan\beta$
are shown in Fig.~\ref{fig:mssmhiggs}. 
The exclusion regions depend only very mildly on assumptions on the sign of
the Higgs mass term $\mu$ and the mixing in the scalar top quark sector.

\section{Searches for supersymmetry}
Supersymmetry (SUSY) is one of the most appealing extensions of the SM, as it
solves the hierarchy problem and could provide a candidate for cold
dark matter. 
Supersymmetric models predict the existence of scalar leptons and quarks
and spin-1/2 gauginos as super-partners of the standard model
leptons, quarks and gauge bosons. 
$R$-parity is introduced as a new
multiplicative quantum
number to differentiate between standard model ($R=1$) and supersymmetric
($R=-1$) particles. As a consequence of the assumption of $R$-parity
conservation, supersymmetric particles are produced in pairs and the lightest
supersymmetric particle (LSP) needs to be stable. In supersymmetric models
inspired by supergravity, the lightest neutralino
$\tilde{\chi}_1^0$, which is a mixture of the super-partners of the neutral
electroweak gauge and Higgs bosons, is usually assumed to be the LSP and is a candidate for
cold dark matter.
In the following only searches for supersymmetry inspired by minimal
supergravity (mSUGRA) and with the assumption of $R$-parity conservation are
presented. Both, the CDF and D0 collaborations performed many searches within
other supersymmetric models.

\subsection{Gaugino pair production}
The associated production of a chargino-neutralino pair,
$\tilde{\chi}_1^\pm\tilde{\chi}_2^0$, can lead to event topologies
with three leptons, which has a low SM background. The third lepton might be
relatively soft, depending on the mSUGRA parameter space.

\begin{figure}
  \begin{center}
    \includegraphics[width=0.41\textwidth,bb=0 0 520 540,clip]{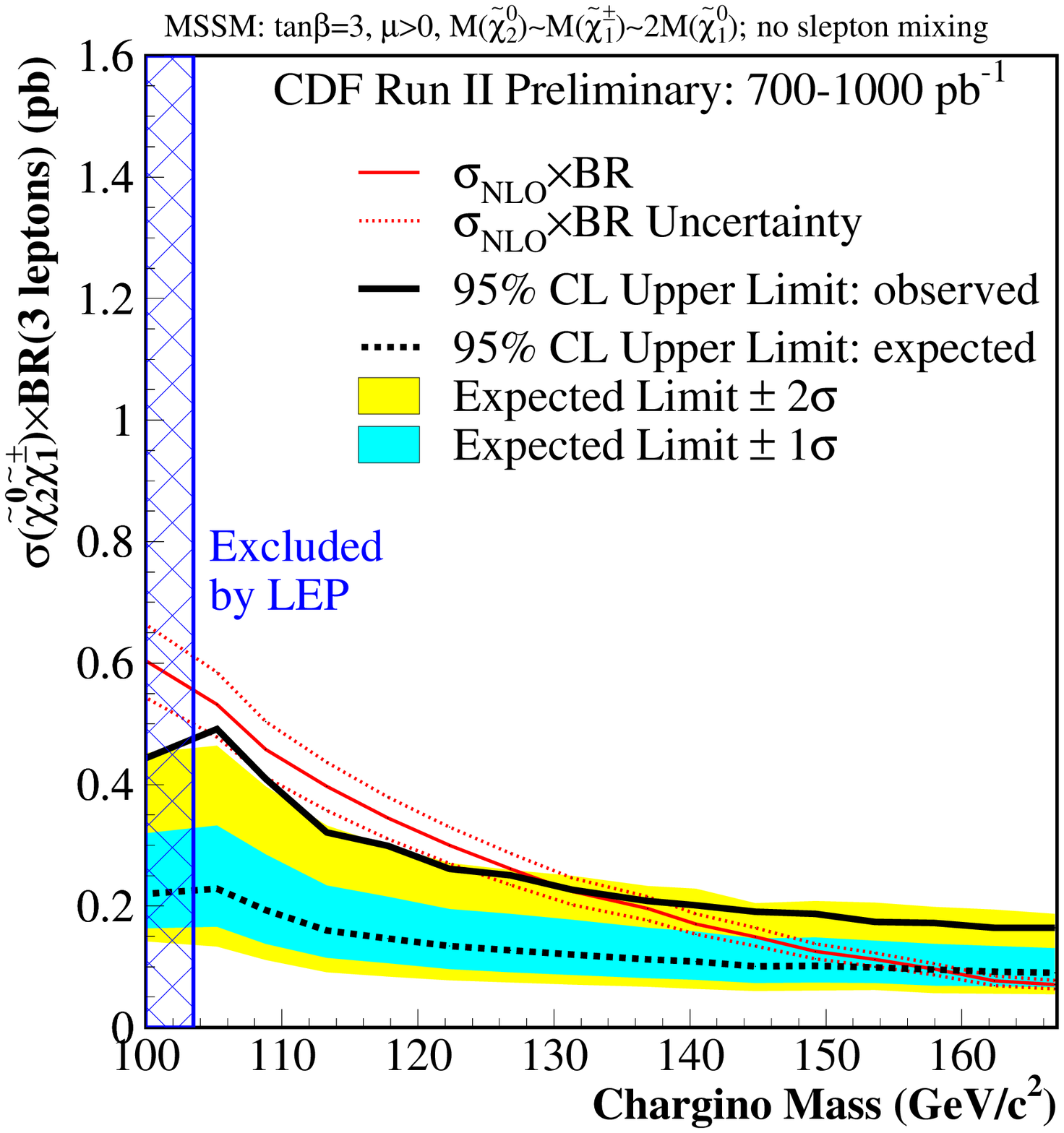}
    \includegraphics[width=0.56\textwidth]{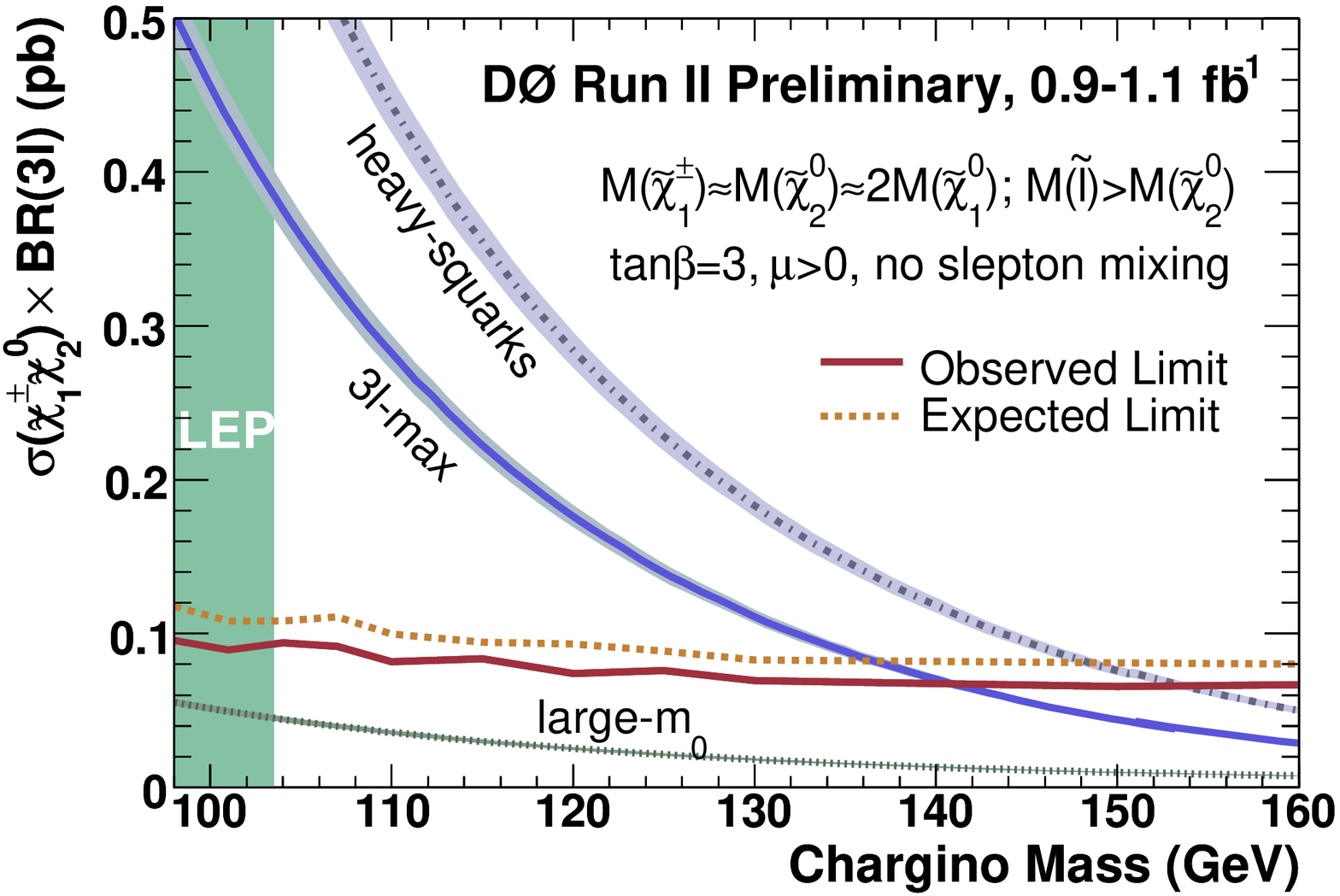}
    \caption{\it Limit on associated $\tilde{\chi}_1^\pm\tilde{\chi}_2^0$
      production in comparison with the expectation of several SUSY scenarios
      obtained by the CDF and D0 collaborations.
    \label{fig:tril}}
  \end{center}
\end{figure}

Both, the CDF and D0 experiments have searched for the tri-lepton
signature taking into account all three lepton flavors and using integrated
luminosities up to 1.1\,\fbi\cite{cdf8653,d05348}. The sensitivity could
be increased by not requiring explicit lepton identification for the third 
lepton and by including final states consisting of a same-sign di-lepton pair.
Both experiments derived limits on the cross-section times branching fraction,
shown in Fig.~\ref{fig:tril}, which are compared to different mSUGRA inspired
scenarios to obtain lower bounds on the chargino mass.

\subsection{Scalar quark and gluino production}
If sufficiently light, squarks and gluinos could be produced in pairs at the
Tevatron. If $M(\tilde{q})<M(\tilde{g})$, mostly pairs of squarks would be
produced, which decay via $\tilde{q}\ra q\tilde{\chi}_1^0$, resulting in an
event signature of two acoplanar jets and \MET{}.
If $M(\tilde{g})>M(\tilde{q})$, gluinos would decay according to
$\tilde{g}\ra q\bar{q}\tilde{\chi}_1^0$ and their pair-production would
give topologies with many jets and \MET{}. In the case of
$M(\tilde{g})\approx M(\tilde{q})$ and $\tilde{q}\tilde{g}$-production the
final state is expected to often consist of three jets and \MET{}.

\begin{figure}
  \begin{center}
    \includegraphics[width=0.6\textwidth,bb= 20 25 540 525,clip]{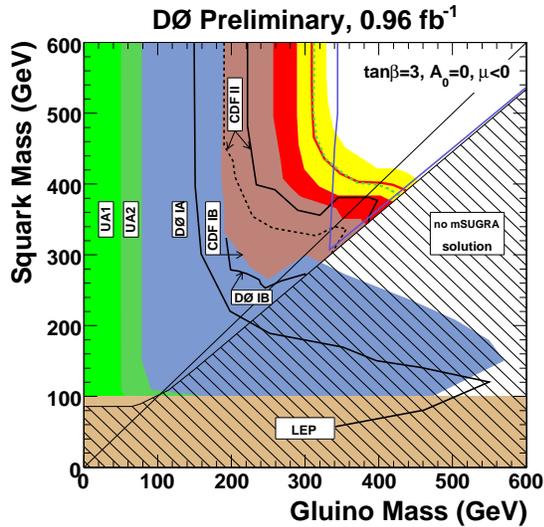}
  \caption{\it Exclusion region in the squark and gluino mass plane.
    \label{fig:sqg}}
  \end{center}
\end{figure}

The D0 collaboration searched for the production of squarks and gluinos
using three different event selections which were targeted at the scenarios
described above\cite{d05312}. The exclusion region in the plane
given by the squark and gluino masses is shown in Fig.~\ref{fig:sqg}.
For the most conservative assumptions (and for $\tan\beta=3$, $A_0=0$, $\mu<0$)
squark and gluino mass limits of $m_{\tilde{q}}>375\,\gev$ and  
$m_{\tilde{g}}>289\,\gev$, respectively, were derived. 
When interpreting the cross-section
limits within mSUGRA the constraints on the common scalar and gaugino masses at
the unification scale, $m_0$ and $m_{1/2}$, could be improved with respect to
limits from LEP.

\subsection{Scalar top and bottom quark production}
Due to a possible large mixing between the super-partners of the left and right
handed top (bottom) quarks, the lighter eigenstate of the scalar top (bottom)
quark might be significantly lighter than the super-partners of the 
other quarks.
Both experiments searched for the pair production of scalar bottom and scalar
top quarks\cite{cdf8411,Abazov:2006fe,Abazov:2006wb}.
The scalar bottom quarks were assumed to decay via 
$\tilde{b}\ra b\tilde{\chi}^0_1$ and the scalar top quarks via the loop
induced decay $\tilde{t} \rightarrow c \tilde{\chi}^0_1$. 
Exclusion regions in the plane given by the sbottom (stop) and neutralino 
masses
were derived reaching $m_{\tilde{b}}\approx 220\,\gev$ and  
$m_{\tilde{t}}\approx 130\,\gev$, respectively.

\section{Conclusions and Perspectives}
The CDF and D0 experiments at the Tevatron collider have performed a
multitude of searches for the standard model and supersymmetric Higgs
boson as well as for signals of supersymmetry. 
At the time of the conference, the searches for the SM Higgs 
boson which include luminosities up to 1\,\fbi{} reached a sensitivity of
a factor 10 (3) times the SM expectation at $m_H\approx 115\,\gev$ 
($m_H\approx 160\,\gev$). Imminent improvements of the limits are expected
from the increased luminosity and refinements in the $b$-tagging and the
event selection.
The ``hint'' of an MSSM Higgs boson at $m_A\approx 160\,\gev$ obtained by CDF
was not confirmed by D0. 
No signal for supersymmetry has yet been found at the Tevatron and stringent
limits, which are significantly improved compared to Run\,I, were set.
At the beginning of 2007 both experiments have recorded integrated 
luminosities exceeding 2\,\fbi{} and are expected to collect much larger data
sets during the full period of Run\,II.
Thus, the sensitivity to the production of the Higgs boson and supersymmetric
particles will substantially improve in the following years.

\section*{Acknowledgments}
I would like to thank my colleagues from the CDF and D0 collaborations for
providing their excellent results and the organizers for the stimulating
conference.

\end{document}